\documentclass[12pt,preprint]{aastex}
\usepackage{rotating}

\newcommand{\kms}{km s$^{-1}$}

\slugcomment{Submitted to The Astrophysical Journal}

\def\ltsima{$\; \buildrel < \over \sim \;$}
\def\simlt{\lower.5ex\hbox{\ltsima}}
\def\gtsima{$\; \buildrel > \over \sim \;$}
\def\simgt{\lower.5ex\hbox{\gtsima}}
\def\Msol{M$_\odot$}

\begin{document}

\title{\bf How Hot is the Wind from TW Hydrae?\altaffilmark{1}}

\author{Christopher M. Johns--Krull}
\affil{Department of Physics \& Astronomy, Rice University, 6100 Main St.
       MS-108, Houston, TX 77005}
\email{cmj@rice.edu}

\author{Gregory J. Herczeg}
\affil{California Institute of Technology, MC105-24, 1200 East California
       Boulevard, Pasadena, CA 91125}
\email{gregoryh@astro.caltech.edu}

\altaffiltext{1}{Based on observations with the NASA/ESA Hubble Space 
Telescope, obtained at the Space Telescope Science Institute, which is 
operated by the Association of Universities for Research in Astronomy, Inc.
under NASA contract NAS5-26555.  These observations are associated with 
program \#GTO-7718.  Also based on bservations made with the
NASA-CNES-CSA Far Ultraviolet Spectroscopic Explorer, which is operated for NASA by the Johns Hopkins University under NASA contract NAS5-32985.}

\begin{abstract} 

It has recently been suggested that the winds from Classical T Tauri
stars in general, and the wind from TW Hya in particular, reaches
temperatures of at least 300,000 K while maintaing a mass loss rate
of $\sim 10^{-11}$ \Msol yr$^{-1}$ or larger.  If confirmed, this would place
strong new requirements on wind launching and heating models.  We
therefore re-examine spectra from the Space Telescope Imaging 
Spectrograph aboard the Hubble Space Telescope and spectra from the
Far Ultraviolet Spectroscopic Explorer satellite in an effort to
better constrain the maximum temperature in the wind of TW Hya.  We
find clear evidence for a wind in the \ion{C}{2} doublet at 1037 \AA\ and
in the \ion{C}{2} multiplet at 1335 \AA.  We find no wind absorption  in
the \ion{C}{4} 1550 \AA\ doublet observed at the same time as the \ion{C}{2}
1335 \AA\ line or in observations of \ion{O}{6} observed simultaneously
with the \ion{C}{2} 1037 \AA\ line.  The presence or absence of \ion{C}{3} 
wind absorption is ambiguous.  The clear lack of a wind in the \ion{C}{4}
line argues that the wind from TW Hya does not reach the 100,000 K
characteristic formation temperature of this line.  We therefore argue
that the available evidence suggests that the wind from TW Hya, and
probably all classical T Tauri stars, reaches a maximum temperature
in the range of 10,000 -- 30,000 K.

\end{abstract}


\section{Introduction} 

Mass loss in the form of a wide-angle wind and/or collimated jet is a
common feature observed in young, low mass stars that are still 
surrounded by accretion disks (for a review see Bally, Reipurth, and
Davis 2006).  The material in these disks eventually accretes onto the 
central star, gets ejected in an outflow, or is incorporated into planets 
or other solar sytem-like bodies.  Understanding the processes through 
which young stars interact with and eventually disperse their disks is 
critical for understanding the rotational evolution of stars and the 
formation of planets.  Mass loss appears to be a necessary component in 
the accretion process both observationally as mentioned above and 
theoretically: angular momentum must be carried away in order to allow 
disk and stellar accretion to occur, and magnetized winds are efficient 
at removing angular momentum (e.g. Pudritz, Pelletier, and Gomez de
Castro 1991, Shu et al. 1994).  

Signatures of mass loss are observed primarily in optical, infrared, and radio 
wavelength spectral diagnostics.  Bipolar outflows are observed in millimeter
wavelength CO lines  and other molecular tracers in the mm and sub-mm range
(e.g. see review by Arce et al. 2006).   In the infrared, emission in H$_2$
(e.g. Gueth \& Guilloteau 1999; Stanke, McCaughrean, \& Zinnecker 2002) and 
[\ion{Fe}{2}] (e.g.  Reipurth et al. 2000) lines are regularly seen in 
outflows from young stars.  At optical wavelengths, Herbig-Haro (HH) objects
and their associated stellar jets are routinely imaged in narrow band
filters centered on H$\alpha$ or on a host of forbidden lines.  High
resolution images of the central engine show that these jets originate 
from young stars surrounded by circumstellar disks, with the jet usually
emerging perpendicular to the disk such as clearly seen in the case of
HH 30 (Burrows et al. 1996, Ray et al. 1996).  Spectroscopically, winds and
jets are diagnosed by the appearance of P-Cygni like line profiles in
permitted lines such as H$\alpha$ (and other Balmer lines) and Na D 
(e.g. Mundt 1984; Reipurth. Pedrosa,
\& Lago 1996; Alencar \& Basri 2000), and in the profiles of forbidden lines 
which often show a strong asymmetry with substantial blue-shifted 
emission with little or no red-shifted emission (e.g. Edwards et al. 1987,
Hamann 1994).  More recently, jets and winds from CTTSs have been
detected at shorter wavelengths corresponding to higher temperature 
emissions.  Hartigan et al. (1999) detect several ultraviolet (UV)
\ion{Fe}{2} emission lines in the HH 47A bow shock, and several irradiated
jets with UV signatures have been discovered recently (e.g. Reipurth 
et al. 1998, Bally \& Reipurth 2001).  Gomez de Castro and Verdugo (2001)
identify feature in the UV semiforbidden emission lines of \ion{C}{3}]
$\lambda1908$ and \ion{Si}{3}] $\lambda1892$ which they associate with shocks
at the base of the jet in a few CTTSs.  At X-ray energies, a number of
jets have also been recently detected (e.g. Pravdo et al. 2001;
Favata et al. 2002; Bally, Feigelson, \& Reipurth 2003).  This X-ray emission
forms in the fastest shocks associated with these stellar jets (see also Bally
et al. 2006).  While these high excitation emissions appear associated 
with shocks within the jets from young stars, until recently it was generally
thought that the stellar or disk wind at the base of these jets was only
heated to a level which could produce features such as the the P-Cygni like
profiles visible in the Balmer lines.  

     A hot wind component, possibly with an origin on the star as opposed
to in the disk, has recently been proposed by a few authors.  Beristain,
Edwards, and Kwan (2001) argue for a hot wind component in CTTSs based on
the analysis of high resolution \ion{He}{1} and \ion{He}{2} line profiles
observed in the optical.  Takami et al. (2002) and Edwards et al. (2003) find
obvious blue-shifted absorption below the local continuum in the 
\ion{He}{1} $\lambda$10830 line in 6 young, low mass stars, again providing
evidence for a hot wind component forming relatively near the star.
Edwards et al. (2003) point out that the strength and ubiquity of this
absorption indicates that the wind emanates over a large solid angle from 
the star: the \ion{He}{1} $\lambda$10830 line does not trace a highly
collimated flow.  These results are reinforced by Edwards et al.
(2006), who found subcontinuum \ion{He}{1} $\lambda$10830 wind absorption in 26 of 38 CTTSs.
%
%
The $\lambda5876$ and $\lambda10830$ lines of 
\ion{He}{1} require a strong ionizing flux or a high temperature for 
excitation, as their upper states lie $\sim20$ eV above the ground level.
If these lines are primarily collisionally excited, temperatures of 
$\sim 25,000$ K are required to excite them (Athay 1965; Avrett, Vernazza, \&
Linsky 1976).  With photoionization followed by recombination and cascade, 
helium excitation can take place at local kinetic temperatures between 8000 
and 15,000 K (Zirin 1975; Heasley, Mihalas, \& Poland 1974; Wahlstrom \& 
Carlsson 1994).   CTTSs in general (e.g. Feigelson et al. 2002), and 
TW Hya in particular (Kastner et al. 2002), are strong X-ray sources, so
the detection of these \ion{He}{1} lines in CTTS winds still leaves a 
wide range of possible temperatures present in these outflows.   Dupree
et al. (2005) detect \ion{He}{1}  $\lambda10830$ absorption in the wind
from TW Hya and T Tau.  They combine these observations with an analysis
of the line profiles of \ion{C}{3} $\lambda977$ and \ion{O}{6} $\lambda1032$
observed with the Far Ultraviolet Spectroscopic Explorer ({\it FUSE}) satellite
to suggest that CTTSs possess continuous, smoothly accelerating winds which 
reach velocities of $\sim 400$ \kms\ and temperatures of $\sim 300,000$ K.
These results are based primarily on the {\it FUSE} data for TW Hya, since 
the data for T Tau is much lower in quality.  Dupree et al. (2005) find a
minimum mass loss rate for the \ion{O}{6} line of $2.3 \times 10^{-11}$
\Msol yr$^{-1}$.

     Such a hot temperature for the wind from TW Hya is surprising, 
particularly given the high mass loss rate.  Alencar \& Batalha (2002) and
Herczeg et al. (2004) estimate a mass accretion rate onto TW Hya of 
$\sim2 \times 10^{-9}$ \Msol\ yr$^{-1}$.  Most estimates suggest that the 
mass loss rate from CTTSs is 0.1 -- 0.3 times that of the mass accretion 
rate (K\"onigl \& Pudritz 2000, Shu et al.  1994), suggesting the mass loss
rate from TW Hya may be as high as $5\times 10^{-10}$ \Msol\ yr$^{-1}$.  
Such a high temperature, high mass loss rate
wind puts strong constraints on the ultimate origin and heating of winds
from CTTSs.  Current theories of mass loss from CTTSs primarily produce
cold, magnetocentrifugally driven flows from the disks around these stars
(e.g. K\"onigl \& Pudritz 2000, Shu et al. 2000) which are then heated
by a combination of ambipolar diffusion and X-rays to temperatures of
$\sim 10^4$ K (Shang et al. 2002; Shang, Li, \& Hirano 2006).  Photoevaporation
of the surface layer of the disk may also produce a low velocity ($\sim 10$ 
\kms) wind with a characteristic temperature of $10^4$ K as well (e.g.
Matsuyama, Johnstone, \& Hartmann 2003; Font et al. 2004).  Thus, if the
Dupree et al. (2005) suggestion holds true, a totally new heating, and
possibly driving, mechanism must be identified.  The analysis presented
by Dupree et al. (2005) relies on the asymmetric {\it shape} of the line
profiles tracing the hot gas: no true absorption against a local continuum
is detected.  Therefore, we re-examine the published (Herczeg et al. 2002, 
2004; Dupree et al. 2005) high resolution spectra from the Space Telescope 
Imaging Spectrograph ({\it STIS}) aboard the Hubble Space Telescope 
({\it HST}) and from the {\it FUSE} satellite to look 
for firm wind signatures and confirm the high proposed temperature in the 
wind of TW Hya.  The {\it STIS} and {\it FUSE} bandpasses cover several lines
with characteristic formation temperatures ranging from $\sim 10^4$ K to 
$\sim 300,000$ K.  We find firm evidence against the proposed wind in the 
\ion{C}{4} and \ion{O}{6} lines of TW Hya.  We therefore conclude that the 
wind in fact does not reach (at least at large optical depth as claimed) a 
temperature of $\sim 100,000$ K, characteristic of the formation of the 
\ion{C}{4} line.  In \S 2 we describe the observational data.  Section 3 
presents our analysis, and \S 4 gives a discussion of our results.

\section{Observations and Data Reduction}

We observed TW Hya with the E140M echelle grating and
the $0\farcs5\times0\farcs5$ aperture on {\it HST}/STIS for 2.3
ks as part of {\it HST} program GTO-7718.  The spectrum spans from
1170--1700 \AA\ with $R=25,000$.  
The data were reduced using standard
calSTIS reduction package written for {\it IDL}.  The flux calibration is
accurate to $\sim 10$\%.  The wavelength calibration is accurate to 
$\sim 5$ \kms.  This spectrum was described by Herczeg et al. (2002).

We also observed TW Hya with the LWRS ($30^{\prime\prime}$) aperture on
{\it FUSE} for a total of 32.8 ks in programs GO-C067 and GTO-P186. 
The {\it FUSE} consists of four co-aligned telescopes, each of
which have two channels coated with LiF and SiC (Moos et al. 2000).  Each
observation yields eight independent spectra covering $\sim 90$ \AA.  Our
co-added spectrum spans from
905--1187 \AA\ with $R\sim15,000$.

We reduced the {\it FUSE} spectra using version 3.1.3 of the calFUSE standard data
reduction pipeline$^1$.   Pulse height values for each LiF and SiC spectrum
were restricted to those detected in the individual extraction windows to
reduce background noise.  We use the standard calFUSE wavelength solution;
however, the zero-point for this solution is not correct and must be fixed
using the observed spectrum.  We calibrated wavelengths at $\lambda>1000$ \AA\
using 25 different H$_2$ lines in this region, setting the wavelength zero
point so that these lines appear at the radial velocity of
the star (Herczeg et al. 2002).  We calibrated the zero point of wavelengths at 
$\lambda<1000$ \AA\ by cross-correlating the emission profiles for the 
\ion{N}{3} 991.5 \AA\ lines and several \ion{O}{1} airglow lines, which appear
in both the longer (calibrated with the H$_2$ lines) and shorter-wavelength 
channels that overlap between 985--1005 \AA.  The error
in our wavelength solution is $\sim 10$ \kms\ at $\lambda>1000$ \AA\ and
$\sim 15$ \kms\ at $\lambda<1000$ \AA.  The absolute and relative flux
calibration is accurate to $\sim 10\%$ and $\sim 5$\%, respectively, at
wavelengths discussed in this paper.  The background subtraction is
accurate to $0.5-3\times10^{-15}$ erg cm$^{-2}$ s$^{-1}$ \AA$^{-1}$.  To
reduce possible contamination by airglow lines, we restrict our analysis
to data obtained during {\it FUSE} nighttime.
\footnotetext[1]{http://fuse.pha.jhu.edu/analysis/calfuse.html}


\section{Analysis}

\subsection{Wind Absorption of H$_2$ and Below the Local Continuum}

     Dupree et al. (2005) assert the presence of a hot ($\sim 300,000$ K) wind
from TW Hya based on the {\it shape} of \ion{O}{6} and \ion{C}{3} emission line profiles.  
When considering the shape of emission lines, there can be an ambiguity
between self-absorption in some part of the line profile and simply a
lack of emission in this same part of the profile.  The presence of a
wind is much more firmly deduced when absorption is detected against a 
local continuum.  Contrary to the statements in Dupree et al. (2005),
a UV continuum is detected from TW Hya, as is evident in many of the figures
shown Herczeg et al. (2002) based on {\it STIS} data.  This continuum is
clearly visible in Figures 1 and 3.  Also evident in Figure 1 is that the 
flux goes to zero in blueshifted wind absorption profiles for key lines from
both the {\it STIS} and {\it FUSE} spectra.


All of the strong, unblended, relatively low temperature lines
present in the {\it STIS} spectrum of TW Hya, including the \ion{C}{2} 1335
\AA\ lines, the \ion{O}{1} 1305 \AA\ triplet, and the \ion{Si}{2} doublets 
at 1260 and 1530 \AA\, show evidence for wind absorption below the local 
continuum.  In the upper panel of Figure 1, we show that the wind absorption
from the stronger, red members of the \ion{C}{2} $\lambda1335$ lines$^1$ stays
at zero flux out 
to a velocity of $\sim -185$ \kms\ and the flux in the wind of the
weaker blue member stays at zero flux out to a velocity of -165 \kms.  
The lower two panels show line profiles from the {\it FUSE} spectrum 
of TW Hya which show clear absorption below the local continuum.  The
middle panel shows the \ion{C}{2} $\lambda1037$ doublet which is blended
with the \ion{O}{6} $\lambda1038$ line.
Absorption below the local continuum is detected in this \ion{C}{2} doublet
and in \ion{H}{1} Ly$\beta$, which are both adjacent to the \ion{O}{6} lines.
The blue member of both the \ion{C}{2} $\lambda$1037 doublet and the
\ion{C}{2} $\lambda1335$ multiplet suffers wind absorption by the red
members (see also \S 3.2).  To quantify the above, we present in Table 1
continuum measurements near each of the lines of interest discussed here.
The continuum in each region is detected at a very significant level, whereas
in the wind absorption region of \ion{C}{2} $\lambda1335.7$ for example, the
measured flux is $1.2 \pm 0.7 \times 10^{-15}$ ergs cm$^{-2}$ s$^{-1}$ 
\AA$^{-1}$ in the velocity range $-90$ to $-170$ \kms.
\footnotetext[1]{We
note that the \ion{C}{2} $\lambda1335$ feature is a triplet with lines
at 1334.532 \AA, 1335.663 \AA, and 1335.708 \AA; however, the middle member
of this triplet should be quite weak due to its low oscillator strength which
is approximately an order of magnitude lower than that of the 1335.708 \AA\
line (Kurucz \& Bell 1995).  Therefore, in Figure 1 we mark only the positions
of the bluemost and redmost members of the multiplet.}

The top panel of Figure 1 also shows that the H$_2$ $\lambda$1333.797 line
is located at $-165$ \kms\ from the \ion{C}{2} $\lambda$1334.5 line,
within the wind absorption (again, the red member of the multiplet shows
the wind absorption in these lines is strong out to at least $-185$
\kms).
%
Of 140 H$_2$ line fluxes from TW Hya modeled by Herczeg et al. (2004), this H$_2$
line is conspicuous in that 
it is the only line that is not well fit.
The observed flux of
$7.9\pm0.7$ erg cm$^{-2}$ s$^{-1}$
in this line is a factor of 6 below its predicted level.  
Herczeg et al. (2002) determine that the
UV H$_2$ emission lines originate in the disk within 2 AU of the star,
and Figure 1 shows that the H$_2$ emission is subject to wind
absorption.

\subsection{The Proposed Hot Wind}

     Dupree et al. (2005) diagnose the presence of a hot wind from TW Hya
by examining the shape of the \ion{C}{3} $\lambda977$ and \ion{O}{6}
$\lambda1032$ emission lines.  These lines display an asymmetry in their
profiles such that there is more emission on the red side of the profile
than on the blue.  This could represent self-absorption or scattering in
a wind, or an accretion flow that preferentially produces red-shifted
emission.  We show these and the \ion{C}{4} $\lambda$1548 line profiles
in Figure 2.  Also shown is a Gaussian fit
to the right side of the line profile, as was done by Dupree et al.
(2005).  The Gaussian is centered at 0 \kms\ and its width and amplitude
are fit to the red wing of the emission line profile.  The red side of
the \ion{C}{3} and \ion{O}{6} are fairly Gaussian in shape.
The \ion{C}{4} line is not well fit by a Gaussian, but the fitting
illustrates the essential points discussed by Dupree et al. (2005).  They
cite the flux deficit on the blue
side of the profile relative to the Gaussian fit as evidence for the
hot wind.  Dupree et al. (2005) then estimate the maximum velocity in the
wind traced by the different diagnostics by estimating where the fit
rejoins the observations in the far blue wing.  For our data reductions
and fits, we find velocities of $\sim -275$ \kms, $\sim -400$ \kms, and $\sim
-400$ \kms\ for \ion{C}{3}, \ion{O}{6}, and \ion{C}{4}, respectively, which
are similar to the values of $-325$ \kms and $-440$ \kms\ for \ion{C}{3} and \ion{O}{6}
found by Dupree et al. (2005).

\subsection{A Closer Look at the {\it STIS} \ion{C}{4} and \ion{C}{2} Lines}

     The \ion{C}{4} $\lambda1550$ doublet provides three very significant
clues which clearly show that there is no wind absorption in these lines.
First, the profile on the blue side of the \ion{C}{4} does not go below
the local continuum (i.e. to a flux of 0) as is the case for the 
wind absorption lines shown in Figure 1.  One may counter that the optical
depth in the \ion{C}{4} lines is simply lower than in the lines shown in
Figure 1.  The additional points address this concern.

An H$_2$ line at 1547.4 \AA\ is located at a velocity of $-165$ \kms\
relative to \ion{C}{4} $\lambda$1548, which is well within any possible
wind absorption due to \ion{C}{4}.  This H$_2$ line has a measured flux of $35.3 \pm 2.7
\times 10^{-15}$ ergs cm$^{-2}$ s$^{-1}$ and a predicted flux of $33.0
\times 10^{-15}$ ergs cm$^{-2}$ s$^{-1}$ (Herczeg et al. 2004).
The blue wing of the \ion{C}{4} 
1548.19 \AA\ line in the immediate vicinity of the $\lambda1547$ H$_2$
line is reduced by a factor of $\sim 5.5$ relative to the Gaussian fit
at this velocity.  Dupree et al. (2005) interpret this as wind absorption
in the \ion{C}{4} line.  The discussion in \S 3.1 above demonstrates that
the H$_2$ lines form inside the wind and will suffer absorption if a wind
absorber overlaps a given H$_2$ line in wavelength.  As a result, we would 
expect the $\lambda1547$
H$_2$ line to have an observed flux about a factor of $\sim 5.5$ below
that predicted by Herczeg et al. (2004), as is in fact seen for the 
$\lambda1334$ H$_2$ line shortward of \ion{C}{2} $\lambda$1334.5 (see \S
3.1).  Since absolutely no absorption is detected in the $\lambda1547$ H$_2$ line, we conclude there is no
significant wind absorption in \ion{C}{4}.

     The third strike against a hot wind visible in the \ion{C}{4} lines
is illustrated in Figure 3.  Shown in black is the line profile
for the blue components of the \ion{C}{2} $\lambda1335$ (top panel) and the
\ion{C}{4} (middle panel) lines.  Shown in red is the red component of the
respective lines overlayed and scaled to match near the peaks and in
the blue wings.  The two components of the \ion{C}{4} line have almost
identical shape, which is somewhat expected since they trace almost
exactly the same material.  The red wing of the 1548.19 \AA\ line extends to
$\sim -130$ \kms\ relative to the 1550.77 \AA\ line, where any wind absorption
due to the 1550.77 \AA\ line should be quite strong.
Looking at the middle panel of Figure 3,
the observed flux in the 1548.19 \AA\ line is down by a factor or $\sim 5.5$
at this velocity.  The red wing of the 1548.19 \AA\ line shows no such
evidence for absorption due to the ``wind'' traced by the 1550.77 \AA\ line.

This is not at all the case for \ion{C}{2}, where we know there truly is a 
wind present.  By overplotting the \ion{C}{2} lines in the top panel of 
Figure 3, we see that the short wavelength (1334.53 \AA)
member of the multiplet shows dramatic absorption due to the wind from
the long wavelength members of the multiplet.
The different members of the multiplet trace essentially the
same material.  The red profile in the upper panel of Figure 3
show that the emission in these lines should extend to $\sim +300$ \kms;
however, the 1334.53 \AA\ line only extends to $\sim +80$ \kms\ due to 
the strong wind absorption due to the red members of the multiplet.

If these \ion{C}{4} lines (absorption and emission components) form
in a wind, the far red wings are formed by material on the far side of the
star.  TW Hya has a large inner hole in its disk (e.g. Johns--Krull \&
Valenti 2001), so this material would be visible.  If the emission 
component to the line forms in the magnetospheric accretion flow, it forms
very near the star.  Regardless of whether
the emission is produced by the accretion flow or the wind, the light would
pass through the wind on the near
side of TW Hya and should suffer absorption at any wavelength where the
projected velocities from the two lines overlap.  The total lack of 
detectable \ion{C}{4} absorption again leads us to conclude there is no
significant amount of \ion{C}{4} in the wind from TW Hya.

\subsection{A Closer Look at the {\it FUSE} \ion{O}{6} and \ion{C}{3} Lines}

The \ion{O}{6} emission line profiles are similar to the \ion{C}{4}
profiles, with strong redshifted emission extending to $\sim +400$ \kms\ and
much weaker blueshifted emission.  The middle panel of Figure 1
demonstrates that the continuum flux goes to zero in wind absorption
due to the nearby \ion{C}{2} $\lambda$1037 doublet and the 
Ly$\beta$ $\lambda 1025.7$ line.  However, Figures 2 and 3 demonstrate that
we do not detect any subcontinuum absorption produced by the stronger
member of the \ion{O}{6} doublet, at 1031.91 \AA.

The long wavelength member of the \ion{C}{2} $\lambda$1037 doublet occurs
at -172 \kms\ relative to the \ion{O}{6} $\lambda1037.6$ line.  The lower panel of
Fig. 3 overlays two \ion{O}{6} lines, and the scaled \ion{C}{2}
$\lambda1335.7$ is used as an approximation for the \ion{C}{2} $\lambda1037.0$ line
profile.  Any wind in the \ion{O}{6} $\lambda$1037.6 line would absorb most
of the \ion{C}{2} $\lambda$1037.0 line, but the red side of the \ion{C}{2}
line is strong, showing no indication of absorption.
We therefore find an absence of any detectable \ion{O}{6} in
the wind.

Whether the \ion{C}{3} line shows wind absorption is ambiguous.  The
sensitivity of {\it FUSE} at 977 \AA\ is lower than at 1035 \AA,
so the continuum is not as well detected.  The continuum level measured
between -100 and -200 \kms\ from line center is $\sim10^{-14}$ erg
cm$^{-2}$ s$^{-1}$ \AA$^{-1}$, which is higher than the
local continuum of $4.5\times10^{-15}$ erg cm$^{-2}$ s$^{-1}$ \AA$^{-1}$.  
No definite \ion{C}{3} wind absorption is detected,
but we also cannot rule out some wind absorption that is either optically thin or narrower than
that seen in other wind absorption lines.  The non-detection is therefore ambiguous.

Some emission is present at -270
\kms, with a flux 10\% that at the peak of the line profile.  The flux in
\ion{C}{4} and \ion{O}{6} lines drops to 10\% the peak value at -150 \kms,
while \ion{C}{2} lines show no emission shortward of -200 \kms.
Therefore, the emission at -270 \kms\ is probably not \ion{C}{3} emission.  An H$_2$ line
at 976.2 \AA\ may contribute as much as half of the flux at -280 \kms.
We are unable to identify the additional emission as either airglow or
other atomic lines.  However, we are unable to use the presence of emission
at -270 \kms\ to infer the presence of \ion{C}{3} wind absorption.

\section{Discussion}

The hot FUV emission lines from CTTSs are related to accretion, as is seen
in correlations with mass accretion rate (Johns--Krull et
al. 2000; Calvet et al. 2004).  The line profiles are often complex, and
can show significant excess redshifted emission.  Such an asymmetric line
profile is naturally expected for optically thin lines that are produced by 
hot gas in downflowing accretion columns.   Dupree et al. (2005) also suggest
that narrow
\ion{O}{6} profiles in the lower-quality spectrum of T Tau could suggest
wind absorption, but the narrow width is instead caused by strong
absorption in interstellar H$_2$ lines (e.g., Roberge et al. 2001, Walter
et al. 2003).  However, asymmetric or narrow profiles of warm emission
lines, including
\ion{C}{2}-\ion{C}{4}, \ion{O}{6}, \ion{He}{2}, \ion{N}{5}, and \ion{Si}{4}
are
insufficient to infer the presence of a wind from CTTSs.
Detecting wind absorption from FUV spectra of CTTSs requires either measuring subcontinuum
absorption or absorption of nearby atomic or H$_2$ lines.  

We conclusively find that no \ion{C}{4} wind absorption is detected in the
STIS spectrum of TW Hya, based on several independent diagnostics.  
Since \ion{C}{4} and \ion{O}{6} trace gas at
$T=100,000$ and $300,000$ K, respectively, the lack of any detectable
\ion{C}{4} in the wind means that no higher temperature \ion{O}{6} gas will 
be detectable.  We confirm this conclusion by finding an absence of \ion{O}{6}
wind absorption in the {\it FUSE} spectrum of TW Hya.  These findings
emphasize the result that asymmetric line profiles with more red emission
than blue as shown here do not necessarily diagnose the presence of a 
wind.  Indeed, Lamzin et al. (2004) argue that the extended blue wings
of \ion{C}{3} $\lambda977$ \AA, \ion{C}{4} $\lambda1550$ \AA, and 
\ion{O}{6} $\lambda1032$ \AA\ arise in the magnetospheric infall around
TW Hya.

The lack of detectable absorption by \ion{C}{4} and the positive detection
of \ion{C}{2} wind
absorption indicates that the abundance of \ion{C}{4} in the wind
is less than 1\% of that of \ion{C}{2}.  Based on the ionization
equilibrium of C, this upper limit suggests a maximum wind temperature of
$<50,000$ K.  The detection of optically thin wind absorption  in the
\ion{He}{1}
$\lambda$10830 \AA\ requires either some photoionization or
temperatures of $>25,000$ K in the wind.  In either
case, some \ion{C}{3} wind absorption is expected.  Unfortunately the {\it
  FUSE} spectrum of TW Hya at 977 \AA\ is not high enough quality to
clearly detect the presence or absence of \ion{C}{3} in the wind.

In addition to \ion{C}{2} and \ion{H}{1} discussed earlier, wind absorption
from TW Hya is detected in \ion{Al}{2}, \ion{Si}{2}, \ion{Mg}{2},
\ion{O}{1}, and \ion{N}{1}, but is not detected in \ion{C}{1}
or \ion{Si}{1}.  Wind absorption is also prevalent from the CTTS RU Lupi
(Herczeg et al. 2005), which has a much higher mass accretion rate than TW
Hya, and therefore presumably a higher wind mass loss rate.  In addition to
the same wind absorption lines as seen in TW Hya, the higher optical depth
in the wind from RU Lupi allows us to detect \ion{Cl}{1}, many \ion{Fe}{2} lines, and
\ion{Si}{3}.  The similarity in
wind absorption lines from RU Lupi and TW Hya suggests that the ionization
of the wind can be similar for stars
with mass loss rates that likely differ by at least a factor of 10.

The absence of \ion{C}{1} absorption in the wind of both TW Hya and RU Lupi 
star suggests that the temperature of the wind is $>20,000$ K.  However,
the presence of \ion{O}{1}, \ion{Cl}{1}, and \ion{Fe}{2} in the wind
requires gas at $T<20,000$ K.
We therefore speculate that the winds from CTTSs have a kinetic temperature
of $\sim 10,000$ K, and that the ionization state is dominated by
photoionization.  Elements with first ionization potential longward of the
Lyman limit are photoionized.  A smaller amount of ionising photons may
also irradiate the wind, which would explain prevalent wind absorption in the
\ion{He}{1} recombination line at 10830 \AA\ (Edwards et al. 2006), 
\ion{N}{1} lines with lower levels excited to 2.4 eV, and the
detection of \ion{Si}{3} in the wind of RU Lupi (Herczeg et al. 2005).  No strong lines occur
between 912--954 \AA, which may explain why \ion{Cl}{1} is not ionized in
the wind of RU Lupi.

\acknowledgements
We would like to acknowledge several useful comments from an anonymous
referee.  CMJ-K acknowledges partial support through program \#AR-9933 
provided by NASA through a grant from the Space Telescope Science Institute, 
which is operated by the Association of Universities for Research in 
Astronomy, Inc., under NASA contract NAS5-26555.  GJH acknowledges partial 
support from {\it FUSE} program GO-C067, provided by NASA contract NAS5-32985.

\clearpage

\begin{figure}
\epsscale{0.7}
\plotone{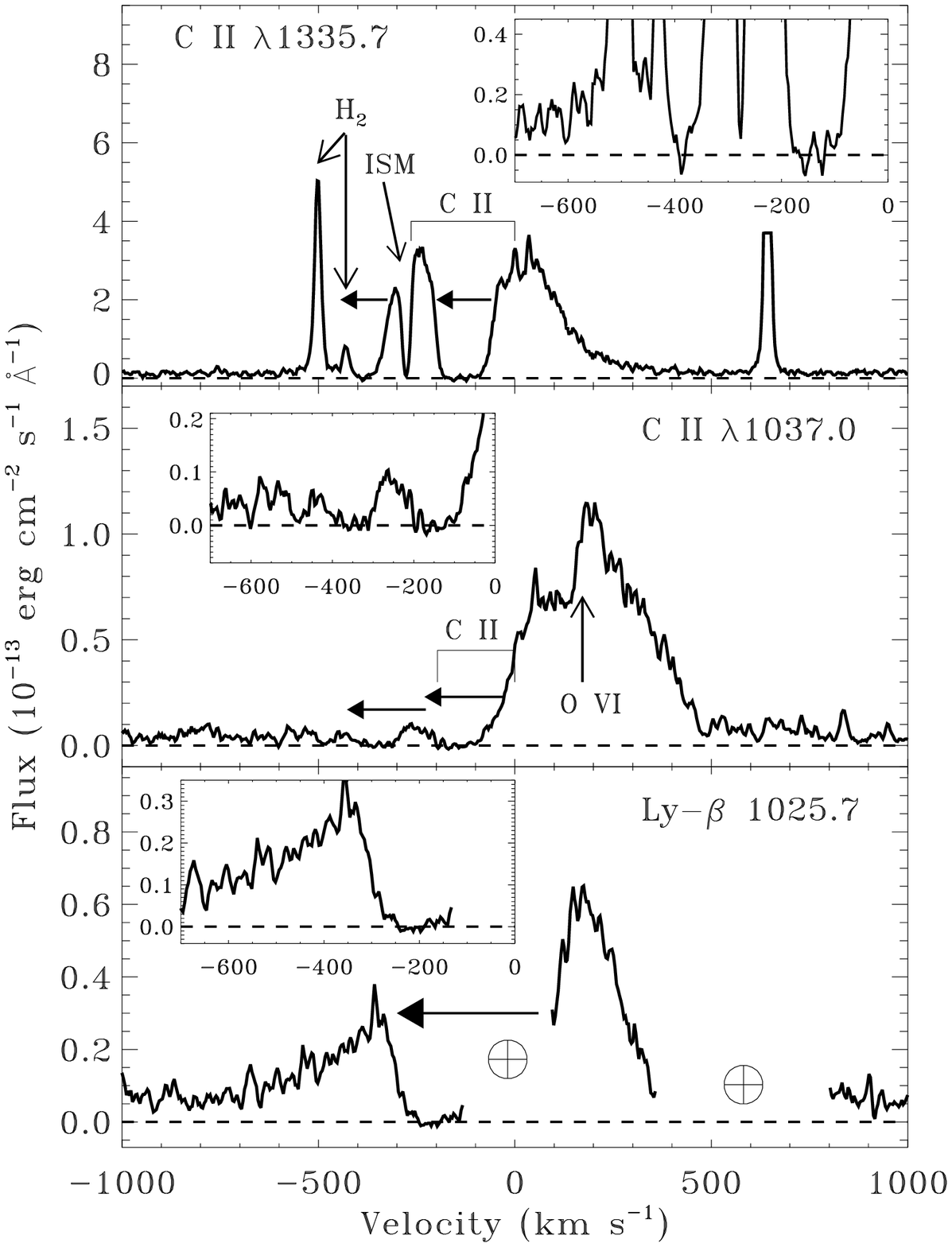}
\caption{Line profiles of UV emission lines in TW Hya which show 
blue-shifted wind absorption below the local continuum.  The \ion{C}{2} 
$\lambda1335$ lines in the upper panel come from the {\it STIS} spectrum.  
The lower two panels show \ion{C}{2} $\lambda1037$, \ion{O}{6}
$\lambda1038$, and Ly-$\beta$ lines in TW Hya observed by {\it FUSE}.  The
dashed horizontal line marks the zero flux level.  Horizontal
arrows indicate the range over which wind absorption is detected.  The 
inserts clearly show this wind absorption below the local continuum in 
each line.}
\end{figure}

\begin{figure} 
\epsscale{0.7}
\plotone{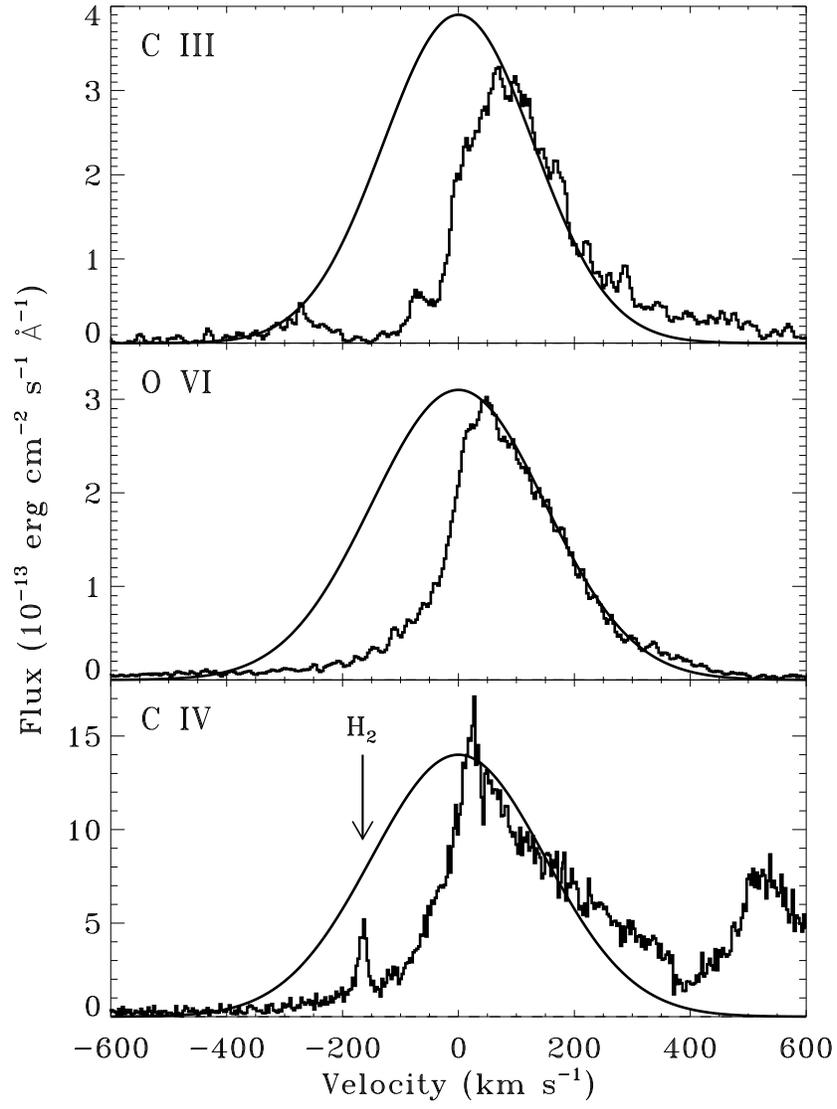}
\caption{Line profiles of \ion{C}{3} $\lambda977$, \ion{O}{6}
$\lambda1032$, and \ion{C}{4} $\lambda1548$.  Shown with each line is
a Gaussian fit to the red side of the profile.}
\end{figure}

\begin{figure} 
\epsscale{0.7}
\plotone{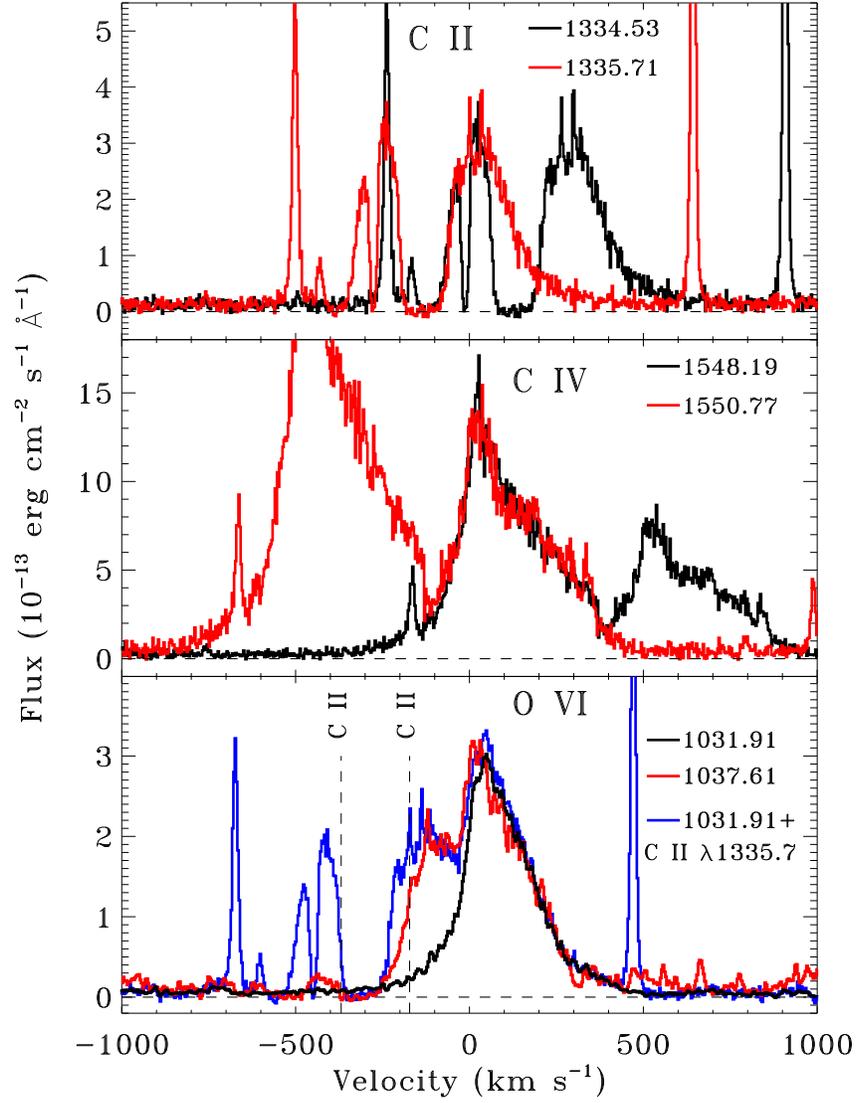}
\caption{Line profiles of \ion{C}{2} $\lambda1335$, \ion{C}{4}
$\lambda1550$, and \ion{O}{6} $\lambda$1035.  The black line shows the blue member of each multiplet.
The red line shows the red member of each multiple scaled to match the
peak and blue wing of the blue member.   The
dashed horizontal line marks the zero flux level.  In the bottom
panel, the blue line shows \ion{C}{2} emission, scaled from the \ion{C}{2}
$\lambda 1335.7$ line, added to the scaled \ion{O}{6} $\lambda1032$ line to
estimate the profile of the \ion{C}{2} $\lambda1037$ line.}
\end{figure}

\clearpage

\begin{table}[t]
\label{tab:obsx.tab}
\caption{Continuum measurements near potential wind absorption features}
\begin{tabular}{cccc}
\hline
Ion & $\lambda$ (\AA) &  Continuum Regions$^a$  (\AA) & Continuum Flux$^b$ \\
\hline
\ion{C}{3} & 977.0 & 973--975.5,979--985 & $3.8\pm0.36$ \\
\ion{H}{1} Ly$\beta$ & 1025.7 &1018--1023,1029.0--1030.7 & $4.93\pm0.19$ \\
\ion{O}{6} & 1031.9 & 1029.0--1030.7,1034.5--1035.5& $5.05\pm0.30$ \\
\ion{C}{2} & 1037.0 & 1034.5--1035.5,1042--1046.3 & $3.82\pm0.17$  \\
\ion{O}{6} & 1037.6 & 1034.5--1035.5,1042--1046.3 & $3.82\pm0.17$  \\
\ion{C}{2} & 1335.0 & 1330--1333,1339-1342 & $13.2\pm0.3$ \\
\ion{C}{4} & 1549.0 & 1544.5--1546 & $23.8\pm0.08$  \\
\hline
\multicolumn{4}{l}{$^a$Regions used to estimate the continuum flux at the respective emission line.}\\ 
\multicolumn{4}{l}{$^b$$10^{-15}$ ergs cm$^{-2}$ s$^{-1}$ \AA$^{-1}$}\\
\end{tabular}
\end{table}


\clearpage

\begin{references}
\reference{} Alencar, S.~H.~P., \& Basri, G.\ 2000, \aj, 119, 1881 
\reference{} Alencar, S.~H.~P., \& Batalha, C.\ 2002, \apj, 571, 378
\reference{} Arce, H.~G., Shepherd, D., Gueth, F., Lee, C.-F., Bachiller,
   R., Rosen, A., \& Beuther, H. 2006, Protostars and Planets V, B. Reipurth,
   D. Jewitt, \& K Keil (eds.), (Tucson: Univ. of Arizona Press), in press
\reference{} Athay, R.~G.\ 1965, \apj, 142, 755 
\reference{} Avrett, E.~H., Vernazza, J.~E., \& Linsky, J.~L.\ 1976, \apjl, 
   207, L199 
\reference{} Bally, J., Feigelson, E., \& Reipurth, B.\ 2003, \apj, 584, 843
\reference{} Bally, J., \& Reipurth, B.\ 2001, \apj, 546, 299 
\reference{} Bally, J., Reipurth, B., \& Davis, C.~J.\ 2006, Protostars and
   Planets V, B. Reipurth, D. Jewitt, \& K Keil (eds.), (Tucson: Univ. of
   Arizona Press), in press
\reference{} Beristain, G., Edwards, S., \& Kwan, J.\ 2001, \apj, 551, 1037 
\reference{} Burrows, C.~J., et al.\ 1996, \apj, 473, 437 
\reference{} Calvet, N., Muzerolle, J., Briceno, C., Hernandez, J.,
Hartmann, L., Saucedo, J. L., \& Gordon, K. D. 2004, AJ, 128, 1294
\reference{} Dupree, A.~K., Brickhouse, N.~S., Smith, G.~H., \& Strader, J.\
   2005, \apjl, 625, L131
\reference{} Edwards, D., Cabrit, S., Strom, S.~E., Heyer, I., Strom, K.~S.,
   \& Anerdson, E.\ 1987, \apj, 321, 473
\reference{} Edwards, S., Fischer, W., Hillenbrand, L., \& Kwan, J.\ 2006,
   \apj, 646, 319
\reference{} Edwards, S., Fischer, W., Kwan, J., Hillenbrand, L., \&
   Dupree, A.~K.\ 2003, \apj, 599, L41
\reference{} Favata, F., Fridlund, C.~V.~M., Micela, G., Sciortino, S., \&
   Kaas, A.~A.\ 2002, \aap, 386, 204
\reference{} Feigelson, E.~D., Broos, P., Gaffney, J.~A., III, Garmire, G., 
   Hillenbrand, L.~A., Pravdo, S.~H., Townsley, L., \& Tsuboi, Y.\ 2002, \apj,
   574, 258 
\reference{} Font, A.~S., McCarthy, I.~G., Johnstone, D., \& Ballantyne, 
   D.~R.\ 2004, \apj, 607, 890
\reference{} G{\'o}mez de Castro, A.~I., \& Verdugo, E.\ 2001, \apj, 548, 976 
\reference{} Gueth, F., \& Guilloteau, S.\ 1999, \aap, 343, 571 
\reference{} Hamann, F.\ 1994, \apjs, 93, 485
\reference{} Hartigan, P., Morse, J.~A., Tumlinson, J., Raymond, J., \& 
   Heathcote, S.\ 1999, \apj, 512, 901 
\reference{} Heasley, J.~N., Mihalas, D., \& Poland, A.~I.\ 1974, \apj, 192,
   181 
\reference{} Herczeg, G.~J., Linsky, J.~L., Valenti, J.~A., Johns--Krull, 
   C.~M., \& Wood, B.~E.\ 2002, \apj, 572, 310 
\reference{} Herczeg, G.~J., Wood, B.~E., Linsky, J.~L., Valenti, J.~A., \& 
   Johns--Krull, C.~M.\ 2004, \apj, 607, 369 
\reference{} Herczeg, G.~J., et al.  2005, \aj, 129, 2777
\reference{} Johns--Krull, C.~M. and Valenti, J.~A. 2001, \apj, 561, 1060
\reference{} Johns-Krull, C. M., Valenti, J. A., \& Linsky, J. L. 2000, ApJ, 539, 815
\reference{} Kastner, J.~H., Huenemoerder, D.~P., Schulz, N.~S., Canizares,
   C.~R., \& Weintraub, D.~A.\ 2002, \apj, 567, 434
\reference{} K\"onigl, A., \& Pudritz, R.~E.\ 2000, Protostars and Planets IV, 
   759 
\reference{} Kurucz, R.~L. \& Bell, B. 1995, Kurucz CD-ROM No. 23: Atomic 
   Line Data, Cambridge, Mass.: Smithsonian Astrophysical Observatory
\reference{} Lamzin, S.~A., Kravtsova, A.~S., Romanova, M.~M., \& Batalha, C.\
   2004, Astronomy Letters, 30, 413 
\reference{} Matsuyama, I., Johnstone, D., \& Hartmann, L.\ 2003, \apj, 582, 
   893 
\reference{} Moos, H.~W., et al.\ 2000, \apjl, 538, L1
\reference{} Mundt, R.\ 1984, \apj, 280, 749
\reference{} Pravdo, S.~H., Feigelson, E.~D., Garmire, G., Maeda, Y., 
   Tsuboi, Y., \& Bally, J.\ 2001, \nat, 413, 708
\reference{} Pudritz, R.~E., Pelletier, G., \& Gomez de Castro, A.~I.\ 1991,
   NATO ASIC Proc.~342: The Physics of Star Formation and Early Stellar 
   Evolution, 539
\reference{} Ray, T.~P., Mundt, R., Dyson, J.~E., Falle, S.~A.~E.~G., \& Raga,
   A.~C.\ 1996, \apjl, 468, L103
\reference{} Reipurth, B., Bally, J., Fesen, R.~A., \& Devine, D.\ 1998, 
   \nat, 396, 343 
\reference{} Reipurth, B., Pedrosa, A., \& Lago, M.~T.~V.~T.\ 1996, A\&AS,
   120, 229
\reference{} Reipurth, B., Yu, K.~C., Heathcote, S., Bally, J., \& 
   Rodr{\'{\i}}guez, L.~F.\ 2000, \aj, 120, 1449
\reference{} Roberge, A., et al. 2001, ApJ, 551, L97
\reference{} Shang, H., Glassgold, A.~E., Shu, F.~H., \& Lizano, S.\ 2002,
   \apj, 564, 853
\reference{} Shang, H., Li, Z.-Y., \& Hirano, N.\ 2006, Protostars and Planets
   V, B. Reipurth, D. Jewitt, \& K Keil (eds.), (Tucson: Univ. of Arizona 
   Press), in press
\reference{} Shu, F.\ H., Najita, J., Ostriker, E., Wilkin, F., Ruden, S., \&
   Lizano, S. 1994, ApJ, 429, 781
\reference{} Shu, F.~H., Najita, J.~R., Shang, H., \& Li, Z.-Y.\ 2000, 
   Protostars and Planets IV, 789 
\reference{} Stanke, T., McCaughrean, M.~J., \& Zinnecker, H.\ 2002, \aap, 
   392, 239 
\reference{} Takami, M., Chrysostomou, A., Bailey, J., Gledhill, T.~M., 
   Tamura, M., \& Terada, H.\ 2002, \apjl, 568, L53 
\reference{} Wahlstrom, C., \& Carlsson, M.\ 1994, \apj, 433, 417
\reference{} Walter, F. M., et al. 2003, AJ, 126, 3076
\reference{} Zirin, H.\ 1975, \apj, 199, L163
\end{references}
\end{document}